\documentclass[12pt,a4paper]{article}
\usepackage[utf8]{inputenc}
\usepackage[T1]{fontenc}
\usepackage{amsmath}
\usepackage{amsfonts}
\usepackage{amssymb}
\usepackage{hyperref}
\usepackage{graphicx}
\begin{document}
	\date{}
	\title{Spherically symmetric gravitational collapse  in the background of Chaplygin gas}
	\author{Anjali Pandey \footnote{anjalipandey903@gmail.com} Rajesh Kumar \footnote{rkmath09@gmail.com} Sudhir Kumar Srivastava \footnote{sudhirpr66@rediffmail.com}\\
		Department of Mathematics and Statistics, \\
		Deen Dayal Upadhyaya Gorakhpur University, Gorakhpur, INDIA.}
	\maketitle
	\begin{abstract}
		This paper considers the spherically symmetric gravitational collapse in the background of Chaplygin gas as dark energy component. The dark energy is assumed to be generalized and modified Chaplygin gas. The exterior of the star is considered to be Schwarzschild de-sitter/anti-de sitter metric and discusses the junction conditions. We have discussed the singularity formation in both forms of chaplygin gas case and it is obtained that such collapsing fluids favor the formation of black-hole.
	\end{abstract}
	\section{Introduction}
	 The gravitational collapse occurs when an astronomical object contracts as a result of its own gravity. During nuclear fusion in the core of the star when internal pressures are insufficient to counteract the gravitational pull of a  star, it continuously collapses due to its own gravity.  The problem of general relativistic gravitational collapse of stars has drawn interest in theoretical physics for a long time starting with the pioneering work of Oppenheimer and Snyder\cite{aa}. The reason for this interest is simple to understand that amongst the few observable phenomena, the gravitational collapse of massive stars in which the general theory of relativity is predicted to play a pertinent role and has astrophysical significance. The research primarily focused on the singularity formation (Black hole or Naked singularity) in gravitational collapse within the framework of general relativity and other gravitational theories are the interesting problems for the theoretical astrophysicists\cite{bb}-\cite{jj}.
	 
	 \par
	  According to present observations, the universe we live in today is made up of about 71\% dark energy and 15\% dark matter. Many drastically different theories, including the cosmological constant, quintessence, DGP branes, Gauss-Bonnet, and dark energy in brane worlds, have been put forth to explain the nature of dark energy and dark matter \cite{kk}-\cite{tt}. It is also exceedingly challenging for contemporary detectors to pick up on these types of exotic fluids because they don't seems to interact with the ordinary matter particles.\cite{uu}-\cite{vv}
	  
	  \par
	   For the stars made of baryonic matter, besides gravitational interaction, they can also interact with each other by means of strong, weak and electromagnetic forces. Ordinary matter and dark energy are only meant to interact through gravity. Therefore, information about its position, mass, density profiles, etc. can only be ascertained by its gravitational effects. Therefore, 'How dark energy affects the gravitational collapse of a star ?' is  an important question.  It is known that dark energy exerts a repulsive force on its surrounding and this repulsive force may prevent the star from collapse. Some recent works have considered the spherically symmetric star consists of Dark energy and Dark matter and discussed the nature of singularity formation \cite{ww}-\cite{ab}. Since not all massive stars collapse to form black holes but instead may become neutron stars or white dwarfs, it is generally accepted that dark energy may be a significant factor in the collapse of massive stars\cite{zz}-\cite{ac}. Some recent works\cite{ab}, \cite{ad}-\cite{af} have taken into consideration the formation of black holes during gravitational collapse in the presence of dark energy. Therefore, it is vital to introduce these new types of matter distribution in our investigations given the idea of the universe's unusual richness.
	   
	   \par
	    In the present works, we considered the spherically symmetric collapsing system in the background of Chaplygin gas as a component of dark energy.  A Chaplygin gas  is characterized by the presssure which is inversly proportional to the energy density ($p\propto \frac{1}{\rho}$). Generalized Chaplygin gas is given by the equation of state (EoS) 
	$p=-\frac{Z}{\rho^{\alpha}}$,	where $Z$ and $\alpha$ are positive real numbers \cite{al}. The generalized Chaplygin gas model was extended to modified Chaplygin gas with Eos
	$p=Y\rho-\frac{Z}{\rho^{\alpha}}$, 	$Y$ and $Z$ are two free parameters and $\alpha$ is positive constant\cite{ak}.
	The paper is organized as follows: In Sec. 2, we consider  shear- free spherically symmetric inhomogeneous star with finite radius, and develop the general formalism for the collapsing problem. In Sec. 3 and 4, we discuss the gravitational collapse of dark energy in the form of generalized Chaplygin gas and Modified Chaplygin gas in order to study the singularity formation during the collapsing process. Sec.6 contains the discussion and concluding remarks of the paper. 
	
	\section{General formulation of the Gravitational collapsing system : Metric and Field equations}
	
	For collapsing configuration, the interior spacetime metric is considered as
	\begin{equation}\label{eqn:2}
		ds^2=A^2dt^2-C^2dr^2-C^2(d\theta^2+sin^2\theta d\phi^2)
	\end{equation}
	where the metric cofficients $A$ and $C$ are functions of (t,r) and the coordinate is taken as $x^{i}=(t,r,\theta,\phi)$ , $i=0, 1, 2, 3$, Here $C(t,r)$ represent the areal radius of the collapsing system.\\
	Consider a spherically symmetric system bounded by spherical surface $\Sigma$ and  interior fluid distribution is considered to be perfect fluid in the form of chaplygin gas 
	\begin{equation}\label{eqn:1}
		T_{ij}=(p+\rho)v_iv_j-pg_{ij}
	\end{equation}
	\\		
	where $\rho$ and $p$ are the energy density and pressure of the fluid respectively and the vector $v_i$ is the four velocity comoving vector satisfying 
	$$v^i=A^{-1}\delta^i_0$$
	The Einstein's field equations $R_{ij}-\frac{1}{2}Rg_{ij}=-kT_{ij}$ $(\text{where the constant} K=\frac{8\pi G}{c^4})$ for the system (\ref{eqn:1}) and (\ref{eqn:2})  yields the following
	\begin{equation}\label{eqn:3}
		\frac{3}{A^2}\frac{\dot{C}^2}{C^2}+ \frac{1}{C^2}+\frac{1}{C^2}\frac{C'^2}{C^2}-\frac{2}{C^2}\frac{C''}{C}=-K\rho
	\end{equation}	
	
	\begin{equation}\label{eqn:4}
		\frac{\dot{C'}}{\dot{C}}-\frac{A'}{A}-\frac{C'}{C}=0
	\end{equation}
	
	\begin{equation}\label{eqn:5}
		\frac{2}{A^2}\frac{\dot{A}}{A}\frac{\dot{C}}{C}+\frac{C'^2}{C^4}+\frac{2}{C^2}\frac{A'}{A}\frac{C'}{C}-\frac{1}{C^2}-\frac{1}{A^2}\frac{{\dot{C}}^2}{C^2}-\frac{2}{A^2}\frac{\ddot{C}}{C}=-kp
	\end{equation}
	
	\begin{equation}\label{eqn:6}
		\frac{1}{C^2}\frac{A''}{A}-\frac{C'^2}{C^4}+\frac{C''}{C^3}+\frac{2}{A^2}\frac{\dot{A}}{A}\frac{\dot{C}}{C}-\frac{1}{A^2}\frac{\dot{C}^2}{C^2}-\frac{2}{A^2}\frac{\ddot{C}}{C}=-kp
	\end{equation}
	where dot $(.)$ and $(')$ denotes differentiation with respest to $t$ and $r$ respectively.
	\par 
	The Energy conservation equation (Bianchi's identity) $T^{i}_{j;i}=0$ have two non-vanishing equations
	\begin{equation}\label{eqn:7}
		3(p+\rho)\frac{\dot{C}}{C}+\dot{\rho}=0
	\end{equation}
	\begin{equation}\label{eqn:8}
		(p+\rho)\frac{A'}{A}+p'=0
	\end{equation}	
	\\
	The expansion scalar $\Theta$ and acceleration vector $(a^i)$ are
	\begin{equation}\label{eqn:9}
		\Theta= v^i_{;i}=\frac{3\dot{C}}{AC}
	\end{equation}
	\begin{equation}\label{eqn:10}
		a^i= v^i_{;j}v^j=\frac{A'}{AC^2}
	\end{equation}
	To study the dynamical properties of the collapsing system, we define the areal velocity $U$ as the collapsing rate of star
	\begin{equation}\label{eqn:z}
		U=\frac{\dot{C}}{C}
	\end{equation}
	Now in view of equs.(\ref{eqn:4}) and (\ref{eqn:z}) we obtain
	\begin{equation}\label{eqn:11}
		A=\mu(t)U
	\end{equation}
	where $\mu(t)$ is an arbitrary function of integration.\\
	
	The mass function $m(t,r)$ which describes the total mass of collapsing fluid at any instant $(t,r)$ is given by \cite{am}
	\begin{equation}\label{eqn:13}
		m(t,r)=\frac{1}{2}C(1+C_{,\alpha}C_{,\beta} g^{\alpha\beta})=\frac{1}{2}C (1+\frac{\dot{C}^2}{A^2}-\frac{C'^2}{C^2})
	\end{equation}
	In view of equations (\ref{eqn:3}) and (\ref{eqn:5}), one can obtain from equ.(\ref{eqn:13}) that
	\begin{equation}\label{eqn:13a}
		\dot{m}= -4\pi p\dot{C} C^2
	\end{equation}
	\begin{equation}\label{eqn:13b}
		m'=4\pi \rho C' C^2
	\end{equation}

	\section{The exterior metric and the junction condition}	
	
	The fluid distribution is comoving inside the spherical surface $\Sigma$ with metric described by equ.(\ref{eqn:2}). The surface $\Sigma (r=r_{\Sigma})$ of the collapsing star choosen to be timelike 3-space and the intrinsic metric on $\Sigma$ is  
	\begin{equation}\label{eqn:17}
		ds^2= \gamma_{ij} d\xi^i\ d\xi^j= d\tau^2-R_0^2(\tau)(d\theta^2+sin^2\theta d\phi^2)
	\end{equation}
	where $\xi^i$= ($\tau$,$\theta$,$\phi$) is intrinsic coordinate and $R_0(\tau)$ is the intrinsic geometrical radius. Since the present work concern with the gravitational collapse in dark-energy background, then the exterior region of spherical system is considered to be the Schwarzschild-de Sitter/anti-de Sitter metric \cite{ag}
	\begin{equation}\label{eqn:18}
		ds^2_+= \alpha(R)dT^2-\alpha^{-1}(R) dR^2+R^2 d\Omega^2
	\end{equation}
	where $\alpha(R)$ is given by
	\begin{equation}\label{eqn:19}
		\alpha(R)= 1- \frac{2M}{R}\pm\frac{\Lambda R^2}{3}
	\end{equation}
	where $\pm$ sign  is corresponding to the Schwarzschild de-Sitter and Schwarzschild anti de-Sitter metric and $\Lambda$ is cosmological constant describing distribution of dark energy in exterior region,  where the coordinate is taken $x^i_+ =(T,R,\theta,\phi)$. In particular, the metric(\ref{eqn:18}) reduces to the Schwarzschild spacetime for $\Lambda=0$\\
	\par
	For the junction condition, we employed the Israel and Darmois condition\cite{mj}-\cite{iw}  which requires the metric(\ref{eqn:2}) and (\ref{eqn:18}) match smoothly across the boundary $\Sigma$ that is\\
	\par
	The continuity of the first fundamental form over $\Sigma$
	\begin{equation}\label{eqn:20}
		(ds_{-}^2)_{\Sigma}= (ds^2)_{\Sigma}= 	(ds_{+}^2)_{\Sigma}
	\end{equation}
	
	and the continuity of the second fundamental form (the extrinsic curvature) over $\Sigma$
	\begin{equation}\label{eqn:21}
		(K^{-}_{ij})_{\Sigma}=(K^{+}_{ij})_{\Sigma}
	\end{equation}
	where $$K^{\pm}_{ij}= -\eta_k^{\pm}\frac{\partial x_{\pm}^k}{\partial x^i x^j}-\eta^{\pm}_k \Gamma^k_{mn}\frac{\partial x^m}{\partial \xi^i} \frac{\partial x^n}{\partial \xi^j}$$
	and $\eta^{\pm}_k$ represent the extrinsic curvature and normal vector on the surface $\Sigma$ in $x^i_{\pm}$ coordinate respectively.
	\par
	Using the junction condition (\ref{eqn:20}) with the metric (\ref{eqn:2}), (\ref{eqn:17}) and (\ref{eqn:18}), we have
	\begin{equation}\label{eqn:22}
		C(t,r_{\Sigma})\quad \underline{\underline{\Sigma}}\quad R\quad\underline{\underline{\Sigma}}\quad R_0(\tau)
	\end{equation}

	\begin{equation}\label{eqn:24}
		A(t,r_{\Sigma})dt= \biggl[\alpha(R_{\Sigma})-\frac{1}{\alpha(R_{\Sigma})}\biggl(\frac{dR_{\Sigma}}{dT}\biggr)^2\biggr]^{\frac{1}{2}}dT= d\tau
	\end{equation}
	
	The extrinsic curvature for the line elements (\ref{eqn:2}) and (\ref{eqn:18}) have the following non-vanishing components
	\begin{equation}\label{eqn:25}
		K^{-}_{00}=-\frac{AA'}{C}\bigg(\frac{dt}{d\tau}\bigg)^2
	\end{equation}
	\begin{equation}\label{eqn:26}
		K^{-}_{22}= C'
	\end{equation}
	\begin{equation}\label{eqn:27}
		K^{-}_{33}= 	K^{-}_{22} Sin^2\theta
	\end{equation}
	and
	\begin{equation}\label{eqn:28}
		K^{+}_{00}=\frac{dR}{d\tau}\frac{d^2T}{d\tau^2}-\frac{dT}{d\tau}\frac{d^2R}{d\tau^2}+\frac{3\alpha'(R)}{2\alpha(R)}\left(\frac{dR}{d\tau}\right)^2\frac{dT}{d\tau}-\frac{\alpha(R)}{2}\alpha'(R)\left(\frac{dT}{d\tau}\right)^3
	\end{equation}
	\begin{equation}\label{eqn:29}
		K^{+}_{22}= \alpha(R) R\frac{dT}{d\tau}
	\end{equation}
	\begin{equation}\label{eqn:30}
		K^{+}_{33}=	K^{+}_{22} \text{Sin}^2\theta
	\end{equation}
	
	In view of equs.(\ref{eqn:26}) and (\ref{eqn:29}) and using (\ref{eqn:22})-(\ref{eqn:24}) the junction condition (\ref{eqn:21}) gives
	\begin{equation}\label{eqn:31}
		\left(\frac{C'}{C}\right)^2-\left(\frac{\dot{C}}{A}\right)^2=\alpha(R)
	\end{equation}
	
	using equ.(\ref{eqn:13}) in above gives
	\begin{equation}\label{eqn:32}
		m(t,r_{\Sigma}) \quad \underline{\underline{\Sigma}}\quad \frac{R}{2}\bigg(1-\alpha(R)\bigg)
	\end{equation}

	taking use of equ.(\ref{eqn:19}) in equ.(\ref{eqn:32}) we obtain

	\begin{equation}\label{eqn:33}
		m(t,r)\quad \underline{\underline{\Sigma}}\quad M\pm \frac{|\Lambda|}{6}R^3
	\end{equation}
	
	Equ.(\ref{eqn:33}) shows that the mass of the collapsing system is equal to the generalized Schwarzschild mass over $\Sigma$. A positive value of $\Lambda$ has an additive contribution and a negative value of $\Lambda$ has a deductive contribution to the collapsing mass.\\
	
	Now from equs.(\ref{eqn:25}) and (\ref{eqn:28})  by using equs. (\ref{eqn:21})-(\ref{eqn:24}), we have 
	\begin{equation}\label{eqn:34}
		\frac{\dot{C}}{C}\frac{d^2T}{d\tau^2}-\frac{dT}{d\tau}\biggl\{\frac{\ddot{C}}{A^2}-\frac{\dot{C}\dot{A}}{A^3}\biggr\}+\frac{d\alpha}{dR}\bigg(\frac{dT}{d\tau}\bigg)\biggl\{\frac{1}{2}+\frac{1}{\alpha(R)}\bigg(\frac{\dot{C}}{A}\bigg)^2\biggr\}=-\frac{A'}{AC}
	\end{equation}
	taking use of eqn.(\ref{eqn:5}) in above, we obtain
	
	\begin{equation}\label{eqn:35}
		p_{DE}\quad \underline{\underline{\Sigma}}\quad-\Lambda
	\end{equation} 
	
	Thus, the equs.(\ref{eqn:33}) and (\ref{eqn:35}) are the required junction conditions. 
	
	\section{Collapse with dark energy in the form of Generalized Chaplygin Gas}
	
	The generalized chaplygin gas is usually defined as a barotropic perfect fluid with equation of state (Eos)\cite{ak}-\cite{al}
	\begin{equation}\label{eqn:36}
		p=-\frac{Z}{\rho^{\alpha}}
	\end{equation}
	
	where $Z$ and $\alpha$ postive real constants.\\
	
	From equation (\ref{eqn:7}) and (\ref{eqn:36}), we have
	\begin{equation}\label{eqn:37}
		3\frac{\dot{C}}{C}-\frac{\rho^{\alpha}\dot{\rho}}{Z-\rho^{\alpha+1}}
	\end{equation}
	
	on integration, it gives
	\begin{equation}\label{eqn:38}
		\rho=\left[\frac{\nu(r)}{C^{3+3\alpha}}+Z\right]^{\frac{1}{1+\alpha}}
	\end{equation}
	\par
	where $\nu(r)$ is an arbitrary function of $r$.\\
	\par
	In view of equs.(\ref{eqn:11}) and (\ref{eqn:36}), the equ.(\ref{eqn:8}) gives,
	\begin{equation}\label{eqn:39}
		\mu(t)U'+\frac{Z\alpha}{(\rho^{1+\alpha}-Z)}\frac{\rho '}{\rho}=0
	\end{equation}
	
	On	integrating above, we obtain
	\begin{equation}\label{eqn:39a}
		U\left[\frac{z-\rho^{1+\alpha}}{\rho^{1+\alpha}}\right]^{\frac{\alpha}{1+\alpha}}=\lambda(t)
	\end{equation}
	where $\lambda(t)$ is arbitrary.
	\par
	By	eliminating $\rho$ between (\ref{eqn:38}) and (\ref{eqn:39a}), one  obtain
	\begin{equation}\label{eqn:41}
		U=\lambda(t)\left[1+\frac{ZC^{3+3\alpha}}{\nu(r)}\right]^{\frac{1+\alpha}{\alpha}}
	\end{equation}
	In collapsing configuration, the self-gravitating system possesses different stages of evolution that determine spacetime singularities as black hole formation or nacked singularity, where the matter density$(\rho \rightarrow \infty)$. Some works also showed the eternally collapse of stellar system and preventing the formation of any kinds of spacetime singularities\cite{ao}-\cite{ap}.\\
	In present work, we discussed the singularity formation with the collapsing rate $U$ of the system: $U\rightarrow 0$ represents collapse-halt (black hole formation), $U\neq 0$ collapsing continue (eternally-collapse), $U\rightarrow \infty$ undefined collapse (No singularity formation) and $C \rightarrow 0$ is the central-singularity of the system.\\
	
	From the equs.(\ref{eqn:38}) and (\ref{eqn:41}), we have
	\begin{equation}\label{eqn:42}
		\rho=\begin{cases}
			\infty, & \text{as}\quad C\to 0,1+\alpha>0\\
			Z^{\frac{1}{1+\alpha}} & \text{as}\quad C \to 0, 1+\alpha<0
		\end{cases}
	\end{equation}
	\par
	and
	\begin{equation}\label{eqn:43}
		U=	\frac{\dot{C}}{C}=\begin{cases}
			\lambda(t), & \text{as}\quad C\to 0,1+\alpha>0\\
			\infty ,&\text{as} \quad C \to 0, 1+\alpha<0
		\end{cases}
	\end{equation}
	One can observed from equs.(\ref{eqn:42}) and (\ref{eqn:43}) that when $(1+\alpha)>0$, the collapse end as black hole for $\lambda(t)\rightarrow 0$. For $(1+\alpha)<0$, the collapsing system denied any kind of central singularity and collapsing rate is undefined.
	
	\section{Collapse with dark energy in the form of modified chaplygin gas}
	The Equation of state of modified chaplygin gas\cite{ak}
	\begin{equation}\label{eqn:47}
		p=Y\rho-\frac{Z}{\rho^{\alpha}}
	\end{equation}
	\par
	where $Y$,$Z$ are two free parameters and $\alpha$ is positive constant.\\
	
	\par
	Using equ.(\ref{eqn:47}) into equ.(\ref{eqn:7}), we have 
	\begin{equation}\label{eqn:48}
		3\frac{\dot{C}}{C}+\frac{\rho^{\alpha}\dot{\rho}}{(Y+1)\rho^{1+\alpha}-Z}
	\end{equation}
	On integrating with respect to 't' we have
	\begin{equation}\label{eqn:49}
		\rho=\left[\frac{Z}{1+Y}+\frac{1}{1+Y} \frac{\nu(r)}{C^{3(1+\alpha)(1+Y)}}\right]^{\frac{1}{1+\alpha}}
	\end{equation}
	where $\nu(r)$ is integrating function. Also, using equs.(\ref{eqn:47}) and (\ref{eqn:4}) into (\ref{eqn:8}), we have
	\begin{equation}\label{eqn:50}
		\frac{\dot{C'}}{\dot{C}}-\frac{C'}{C}+\left[\frac{(Y\rho^{1+\alpha}+Z\alpha)}{(1+Y)\rho^{1+\alpha}-Z}\right]\frac{\rho '}{\rho}=0
	\end{equation}
	\par
	On integrating w.r.to 'r' and using equ.(\ref{eqn:z}), we obtain
	\begin{equation}\label{eqn:51}
		U=\gamma(t)\left[(1+Y)\rho^{1+\alpha}-Z\right]^{\frac{-(ZY+Y+1)}{Z(1+Y)(1+\alpha)}} \rho^{\frac{1}{Z}}
	\end{equation}
	\par
	Now eliminating $\rho$ between (\ref{eqn:49}) and (\ref{eqn:51}), we obtain
	\begin{equation}\label{eqn:52}
		U=\frac{\gamma(t)}{\nu(r)}\left[\frac{Z}{1+Y}+\frac{\nu(r)}{C^{3(1+\alpha)(1+y)}}\frac{1}{1+Y}\right]^{\frac{\alpha}{(1+\alpha)}}C^{3(Y+\alpha+Y\alpha)}
	\end{equation}
	From equ.(\ref{eqn:49})-(\ref{eqn:52}) one can get $\rho \rightarrow \infty$, $U\rightarrow \infty$, $(C(t,r)\neq 0)$ for different values of $Y$ and $\alpha$. For $Y=0$ the system reduces into the case of generalized Chaplygin gas that is equs.(\ref{eqn:49}) and (\ref{eqn:52}) reduces into equs.(\ref{eqn:38}) and (\ref{eqn:41}) respectively.\\
	For $\alpha=1$, we have from (\ref{eqn:49}) and (\ref{eqn:52}) that
	\begin{equation}\label{eqn:53}
		\rho=\left[\frac{Z}{1+Y}+\frac{1}{1+Y}\frac{\nu(r)}{C^{6(1+Y)}}\right]^{\frac{1}{2}} 
	\end{equation}
	
	\begin{equation}\label{eqn:54}
		U=\frac{\gamma(t)}{\nu(r)}\left[\frac{Z}{1+Y}+\frac{1}{1+Y}\frac{\nu(r)}{C^{6(1+Y)}}\right]^{\frac{1}{2}} C^{3(1+2Y)}
	\end{equation}
	In view of equs.(\ref{eqn:53}) and (\ref{eqn:54}) it can be seen that central singularity is formed when $Y>-\frac{1}{2}$. For $Y=-\frac{1}{2}$ and $-1$, $\rho \rightarrow \infty$ and $U\rightarrow \infty$ as $C \rightarrow 0$. Also for $Y< -1$, $\rho = \sqrt{\frac{Z}{1+Y}}$ ( for $Z<0$) and $U\rightarrow \infty$ as C$\rightarrow$ 0, that is no central singularity formed.\\
	Further from equs. (\ref{eqn:49}) and (\ref{eqn:52}) we have for $C\rightarrow 0$

	\begin{equation}\label{eqn:55}
		\rho \rightarrow	\begin{cases}
			\infty, & \text{for $Y = -\frac{1}{2}$}\\
			\infty, & \text{for $Y = -\frac{1}{3}$}\\
			\infty, & \text{for $Y = -\frac{2}{3}$}\\
			\infty, & \text{for $Y= -1$}\\
			(-2Z)^{\frac{1}{1+\alpha}}, & \text{for $Y = -\frac{3}{2}$ and $Z<0$}\\
			\infty, & \text{for $Y =K\geq$ 0 and $\alpha>0 $},
		\end{cases}
	\end{equation}
	\begin{equation}\label{eqn:56}
		U \rightarrow \begin{cases}
			0, & \text{for $Y = -\frac{1}{2}$ and $\alpha >1$}\\
			0, & \text{for $Y = -\frac{1}{3}$ and $\alpha > \frac{1}{2}$}\\
			0, & \text{for $Y = -\frac{2}{3}$ and $\alpha$ > 2 }\\
			\infty, & \text{for $Y$ = -1 and $\alpha$ > 0 }\\
			\infty, & \text{for $Y = -\frac{3}{2}$ and $\alpha>$1}\\
			\infty, & \text{for $Y =-\frac{1}{2}$ and $\alpha$ = 1}\\
			0, & \text{for $Y = K\geq 0$ and $\alpha$ > 0}
		\end{cases} 
	\end{equation}
	where $K \geq 0$ is any constant.
	
	From above equ.(\ref{eqn:55}) and (\ref{eqn:56}) we see that central singularity is formed when $Y=k\geq 0$, $Y=-\frac{1}{2}(\alpha>1)$, $Y=-\frac{1}{3}(\alpha>\frac{1}{2})$, $Y=-\frac{2}{3}(\alpha>2)$ and there is no central singularity for $Y=-\frac{3}{2}$.

	\section{Discussion and Concluding Remarks}
	The structure formation in the universe with the dark-energy background has been taken considerable interest among theoretical physicists and has been applied for the understanding of astrophysical phenomenon. In recent decade, the study of gravitational collapse and singularity formation (black-hole and naked-singularity) have been carried out in the background of dark-energy and dark-matter. The present paper deals with the gravitational collapse of spherically symmetric inhomogeneous fluid with Chaplygin gas as dark-energy components. We have considered the interior metric as a shear-free spherically symmetric and exterior region of star is described by the Schwarzschild de- Sitter/anti- de Sitter space-time containing cosmological constant $\Lambda$. By applying the Darmois- Israel junction condition for matching of interior $(ds^2_{-})$ and exterior $(ds^2_{+})$ line-element at the boundary hypersurface $\Sigma$, we obtained the the two boundary conditions (\ref{eqn:33}) and (\ref{eqn:35}). Condition (\ref{eqn:33}) represents that the collapsing mass is equal to the $M\pm \frac{1}{6}|\Lambda| R^3$, whereas (\ref{eqn:35}) indicates the DE (Chaplygin-gas) pressure is negative/positive according as $\pm \Lambda$  over $\Sigma$. To analyse the physical process of collapsing star, we considered the equation of state of generalised Chaplygin gas and modified Chaplygin gas as a dark-energy component. We have discuss the collapsing scenario and singularity in both the cases of Chaplygin gas. When the collapsing fluid is in the form of generalised Chaplygin gas we obtained that black-hole is formed when $1+\alpha > 0$ and no singularity stage reach when $1+\alpha <0$.
	\par
	When we consider the collapsing fluid in the form of modified Chaplygin gas, we observed a centeral singularity $C\rightarrow 0$ for $1+Y >0$, where $U\rightarrow 0$ and $\rho \rightarrow \infty$ i.e. collapse halt at this stage. For the value $Y=0$ this case reduces into the case of generalised Chaplygin gas and we have the same results as discussed in section(4). Also we evaluate the energy density $\rho$ and collapsing rate $U$ for different value of $Y$ and $\alpha$ where we observed formation of black hole when $Y=k \geq 0, Y=-\frac{1}{2} (\alpha>1), Y=-\frac{1}{3} (\alpha>\frac{1}{2}), Y=-\frac{2}{3} (\alpha>2) $ and there is no central singularity for $Y= -\frac{3}{2}$. Thus our present results show that the DE (Chaplygin gas) component favour the formation of black-hole and hence it has an effect in the collapsing process and play important role in the singularity-formation.

\end{document}